\documentclass[twocolumn,amsmath,amssymb,
showpacs,
 aps,
 lengthcheck,%
]{revtex4-1}
\usepackage{graphicx}
\usepackage{dcolumn}
\usepackage{bm}
\usepackage{hyperref}
\usepackage{makeidx}

\usepackage[english]{babel} 
\usepackage[latin1]{inputenc}
\usepackage{color}
\usepackage{graphicx}

\usepackage{latexsym}
\usepackage{amsthm}

\makeindex

\newcommand{\cG}{{\cal G}}
\newcommand{\cS}{{\cal S}}
\newcommand{\cN}{{\cal N}}
\newcommand{\Sgn}{\text{Sgn}}
\newcommand{\bart}{\bar{t}}
\newcommand{\cQ}{{\cal Q}}

\begin{document}
 
\title{Minimal Excitations in the Fractional Quantum Hall Regime}

\author{J. Rech$^{1}$, D. Ferraro$^{1}$, T. Jonckheere$^{1}$, L. Vannucci$^{2,3}$, M. Sassetti$^{2,3}$, and T. Martin$^{1}$}
\affiliation{$^1$  Aix Marseille Univ, Universit\'e de Toulon, CNRS, CPT, Marseille, France\\
$^2$ Dipartimento di Fisica, Universit\`a di Genova, Via Dodecaneso 33, 16146, Genova, Italy\\
$^3$ CNR-SPIN, Via Dodecaneso 33, 16146, Genova, Italy}

\date{\today}

\begin{abstract}
We study the minimal excitations of fractional quantum Hall edges, extending the notion of levitons to interacting systems. Using both perturbative and exact calculations, we show that they arise in response to a Lorentzian potential with quantized flux. They carry an integer charge, thus involving several Laughlin quasiparticles, and leave a Poissonian signature in a Hanbury-Brown and Twiss partition noise measurement at low transparency. This makes them readily accessible experimentally, ultimately offering the opportunity to study real-time transport of Abelian and non-Abelian excitations.
\end{abstract}

\maketitle

Because of its potential application to quantum information processing, time-dependent quantum transport in open coherent nanostructures attracts prodigious attention.
Recent years have seen the emergence of several attempts to manipulate elementary charges in quantum conductors~\cite{feve07,hermelin11,mcneil11,dubois13}. This opened the way to the field of electron quantum optics (EQO)~\cite{bocquillon14} characterized by the preparation, manipulation and measurement of single-particle excitations in ballistic conductors.

In this context, levitons -- the time-resolved minimal excitation states of a Fermi sea -- were recently created and detected in two-dimensional electron gas~\cite{dubois13, jullien14}, 20 years after being theoretically proposed~\cite{levitov96, ivanov97, keeling06}. 
These many-body states are characterized by a single particle excited above Fermi level, devoid of accompanying particle-hole pairs~\cite{dubois13b}. The generation of levitons via voltage pulses does not require delicate circuitry and has thus been put forward as a solid candidate for quantum bit applications, in particular the realization of electron flying qubits~\cite{bertoni00,yamamoto12}.

Interaction and quantum fluctuations strongly affect low dimensional systems leading to dramatic effects like spin-charge separation and fractionalization~\cite{pham00,leinaas09, giamarchi03}. These remarkable features were investigated by looking at both time-resolved current~\cite{kamata14,perfetto14,calzona15} and noise measurements~\cite{depicciotto97,saminadayar97,berg09,neder12,inoue14}. 
While the emergence of many-body physics and the inclusion of interactions~\cite{bocquillon13,bocquillon13b,ferraro14,wahl14} was recently addressed in the framework of EQO, a conceptual gap still remains when it comes to generating minimal excitations. This is particularly true when the ground-state is a strongly correlated state, as are the edge channels of a fractional quantum Hall (FQH) system~\cite{tsui99}, a situation which has remained largely unexplored so far for time-dependent drives~\cite{jonckheere05}.
The building blocks of such chiral conductors are no longer electrons but instead anyons, which have a fractional charge and statistics~\cite{stern08}. 
For Laughlin filling factors~\cite{laughlin83}, these anyons are Abelian quasiparticles, but more exotic situations involving non-Abelian anyons~\cite{nayak08} are predicted. 
Our understanding of these nontrivial objects would benefit from being able to excite only few anyons at a time~\cite{ferraro15}, allowing us to study their transport and exchange properties, and to combine them through interferometric setups. This calls for the characterization of minimal excitations in the FQH regime.

In this letter, we study levitons in the edge channels of the fractional quantum Hall regime by analyzing the partition noise at the output of a quantum point contact (QPC). Our results rely on a dual approach combining perturbative and exact calculations of the noise in a Hanbury-Brown and Twiss (HBT)~\cite{hanburybrown56, bocquillon12} configuration.
We also provide results in the time-domain, investigating leviton collisions with Hong-Ou-Mandel (HOM)~\cite{hong87, dubois13} interferometry.

\begin{figure}[tb]
\includegraphics[width=8.6cm]{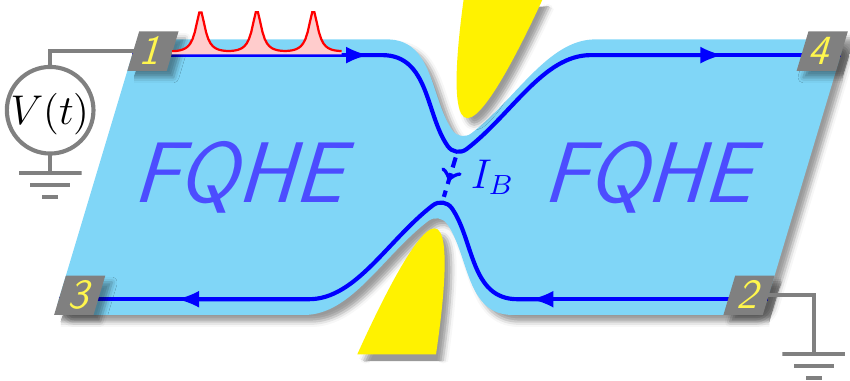}
\caption{Main setup: a quantum Hall bar equipped with a QPC connecting the chiral edge states of the FQH. The left-moving incoming edge is grounded at contact $2$ while the right-moving one is biased at contact $1$ with a time-dependent potential $V(t)$. 
	}
\label{fig:setup}
\end{figure}

Consider a FQH bar (see Fig.~\ref{fig:setup}) with Laughlin filling factor $\nu=1/(2n+1)$  ($n\in\mathbb N$), described in terms of a hydrodynamical model~\cite{wen95} by the Hamiltonian ($\hbar=1$)
\begin{align}
H &= \frac{v_F}{4\pi}  \int dx \left[ 
\sum_{\mu = R,L} \left( \partial_x \phi_\mu \right)^2 
- \frac{2e \sqrt{\nu}}{v_F} V(x,t)~ \partial_x \phi_R 
\right] ,
\end{align}
where the bosonic fields $\phi_{R,L}$ propagate along the edge with velocity $v_{F}$ and are related to the quasiparticle annihilation operator as $\psi_{R,L} (x) = \frac{U_{R,L}}{\sqrt{2\pi a}} e^{\pm i k_F x} e^{-i \sqrt{\nu} \phi_{R,L} (x)}$ (with $a$ a cutoff parameter and $U$ a Klein factor), and $V(x,t)$ is an external potential applied to the upper edge at contact $1$. 

Working out the equation of motion for the field $\phi_R$, $\left(\partial_t +v_F \partial_x\right) \phi_R(x,t) = e \sqrt{\nu} V(x,t)$, one can relate it to the unbiased case using the transformation
\begin{align}
\phi_{R} (x,t) = \phi_{R}^{(0)} (x,t) + e \sqrt{\nu} \int_{-\infty}^{t} dt' V(x',t') ,
\label{eq:newphi}
\end{align}
with $x' = x-v_F(t-t')$, and $\phi_R^{(0)}$ is the free chiral field, $\phi_R^{(0)}(x,t) = \phi_R^{(0)} (x-vt,0)$.
Focusing first on the regime of weak backscattering (WB), the tunneling Hamiltonian describing the scattering between counter-propagating edges at the QPC can be written, in terms of the transformed fields, Eq.~\eqref{eq:newphi}, as $H_T = \Gamma (t) \psi_R^\dagger (0) \psi_L (0) + \text{H.c.}$, where we introduced $\Gamma (t) = \Gamma_0 \exp\left( i e^*  \int_{-\infty}^{t} dt' V(t') \right)$~\footnote{See Supplemental Material for details of the calculation.}, with the bare tunneling constant $\Gamma_0$, the fractional charge $e^* = \nu e$ and assuming a voltage $V(t)$ applied over a long contact, in accordance with the experimental setup~\cite{dubois13}, allowing us to simplify $\int_{-\infty}^{t} dt' V(v_F(t'-t),t') \simeq \int_{-\infty}^{t} dt' V(t')$. 

The applied time-dependent voltage consists of an AC and a DC part $V(t) = V_{dc} + V_{ac} (t)$, where by definition $V_{ac}$ averages to zero over one period $T = 2\pi/\Omega$.  The DC part indicates the amount of charge propagating along the edge due to the drive. The total excited charge $Q$ over one period is then: 
\begin{align}
Q =& \int_0^T dt \langle I (t) \rangle = \nu \frac{e^2}{2 \pi} \int_0^T dt V(t) =q e ,
\end{align}
where the fractional conductance quantum is $G_0 = \nu e^{2}/2\pi$ and the number of electrons per pulse is $q =  \frac{e^* V_{dc}}{\Omega}$. The AC voltage generates the accumulated phase experienced by the quasiparticles $\varphi(t) = e^{*} \int_{-\infty}^t dt' V_{ac}(t')$, characterized by the Fourier components $p_l$ of $e^{-i \varphi(t)}$.

In a 1D Fermi liquid, the number of electron-hole excitations resulting from an applied time-dependent voltage bias is  connected to the current noise created by the pulse scattering on a QPC~\cite{lee93,levitov96,keeling06} which acts as a beam-splitter, as in a HBT setup~\cite{hanburybrown56, bocquillon12}.
For FQH edge states however, scattering at the QPC is strongly non-linear as it is affected by interactions. Special care is thus needed for the treatment of the point contact, and the definition of the excess noise giving access to the number of excitations.

The quantity of interest is the photo-assisted shot noise (PASN), i.e. the zero-frequency current noise measured from contact 3, and defined as 
\begin{align}
\cS &= 2 \int d\tau \int_0^T \frac{d \bart}{T} \langle \delta I_3 (\bart + \frac{\tau}{2}) \delta I_3 (\bart-\frac{\tau}{2}) \rangle
\end{align}
where $\delta I_3 (t) = I_3 (t) - \langle I_3 (t) \rangle$  and the output current $I_3(t)$ reduces, since contact 2 is grounded, to the backscattered current $I_B (t)$, readily obtained from the tunnel Hamiltonian
\begin{align}
	I_B(t) = i e^{*}  \left[ \Gamma (t) \psi_R^\dagger (0,t) \psi_L (0,t)  - \text{H.c.} \right].
\end{align}

When conditions for minimal excitations are achieved in the perturbative regime, excitations should be transmitted independently, leading to Poissonian noise. It is thus natural to characterize minimal excitations as those giving a vanishing excess noise at zero temperature:
\begin{align}
\Delta \cS = \cS - 2 e^{*} \overline{\langle I_B (t) \rangle} ,
\label{eq:DeltqSperturb}
\end{align}
where $\overline{\langle I_B (t) \rangle}$ is the backscattered current averaged over one period.

\begin{figure}[tb]
\includegraphics[width=8.6cm]{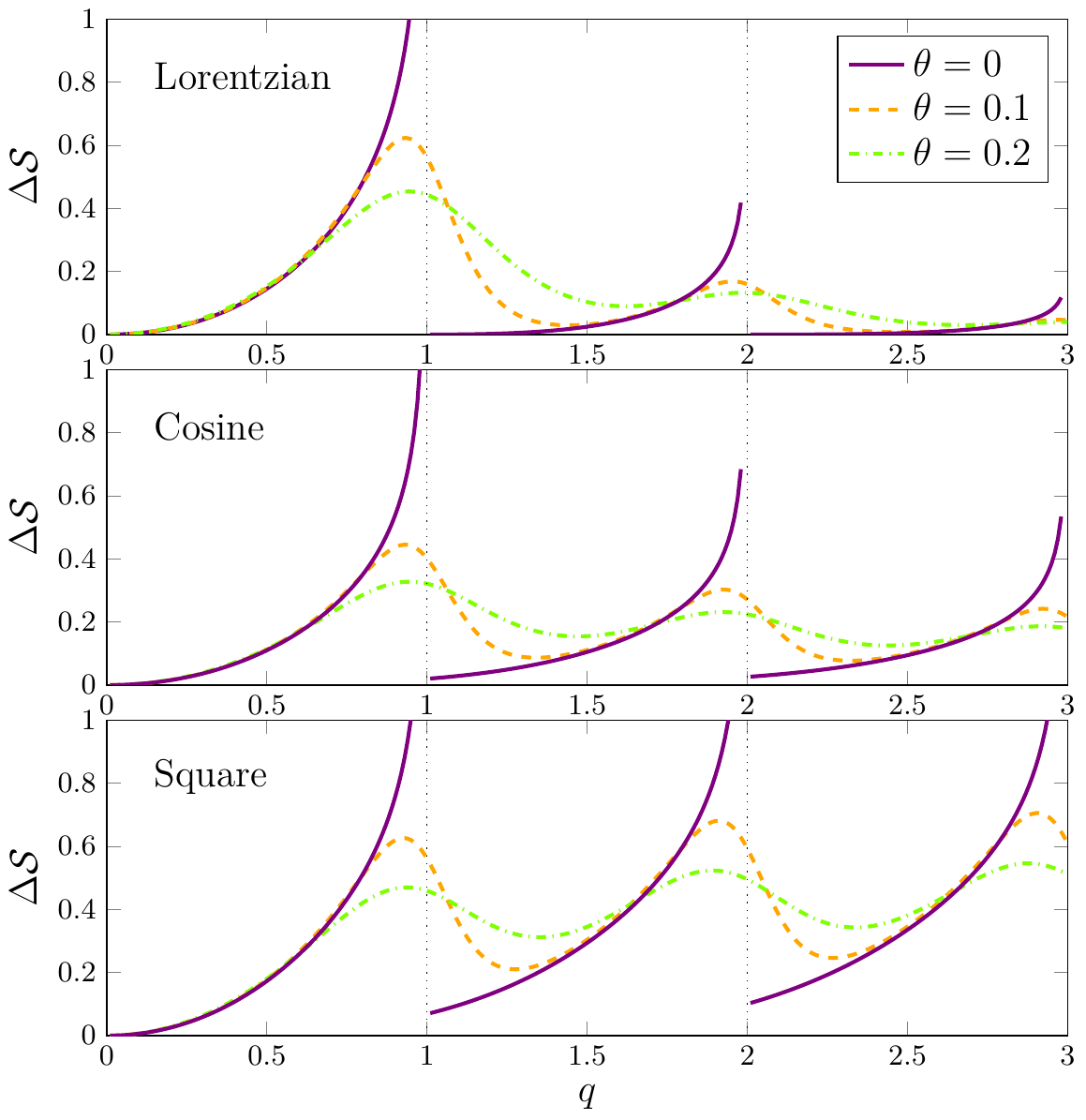}
\caption{Excess noise in units of $S^0 = \frac{2}{T} \left( \frac{e^{*} \Gamma_0}{v_F} \right)^2 \left( \frac{\Omega}{\Lambda} \right)^{2\nu-2}$ as a function of the number of electron per pulse $q$, for different reduced temperatures $\theta$ and filling factor $\nu=1/3$, in the case of a square (bottom), a cosine (middle) and a periodic Lorentzian drive with half-width at half-maximum $\eta=W/T = 0.1$ (top).}
\label{fig:xsnoise}
\end{figure}

Using the zero-temperature bosonic correlation function $\langle \phi_{R/L} (\tau) \phi_{R/L} (0)\rangle_c = - \log \left( 1 + i \Lambda \tau \right)$, this excess noise is computed perturbatively up to order $\Gamma_0^2$, yielding~\cite{Note1}
\begin{align}
\Delta \cS &= \frac{2}{T} \left( \frac{e^* \Gamma_0}{v_F} \right)^2
 \left(\frac{\Omega}{\Lambda}\right)^{2\nu-2} \frac{1}{\Gamma (2 \nu)}
 \nonumber \\
& \quad \times \sum_{l} P_l  \left|  l + q \right|^{2\nu-1} \left[ 1 - \Sgn \left( l + q \right) \right] ,
\end{align}
where $\Lambda = v_F/a$ is a high-energy cutoff and $P_l = |p_l|^2$ is the probability for a quasiparticle to absorb ($l>0$) or emit ($l<0$) $l$ photons, which depends on the considered drive~\cite{Note1}. 
These probabilities $P_l$ also depend on $q$, as the AC and DC components of the voltage are
not independent. Indeed, we are interested here in a periodic voltage $V(t)$ consisting
of a series of identical pulses, with $V(t)$ close to 0 near the beginning and the end of each 
period.  This implies that the AC amplitude is close to the DC one. Our formalism could also be used
to perform a more general analysis  by changing these contributions independently. 
In particular, fixing the DC voltage and changing the AC amplitude allows us to perform a spectroscopy of the probabilities themselves. Conversely, changing the DC voltage at fixed AC amplitudes, we can reconstruct the tunneling rate associated with each photo-assisted process~\cite{dubois13b} in the same spirit as finite frequency noise calculations~\cite{ferraro14b}. 
However, this broader phenomenology does not provide any additional information
concerning the possibility of creating minimal excitation by applying periodic pulses.

In Fig.~\ref{fig:xsnoise}, we show the variation of the excess noise as a function of $q$, for several external drives at $\nu=1/3$ and various reduced temperatures $\theta= k_{B} \Theta/\Omega$ ($\Theta$ the electronic temperature).
At $\theta=0$, only the periodic Lorentzian drive leads to a vanishing excess noise, and only for integer values of $q$. 
This confirms 
that as in the 1D Fermi liquid, and as mentioned in earlier work~\cite{keeling06}, optimal pulses have a quantized flux and correspond to Lorentzians of area $\int dt V = m 2\pi/e^{*}$ (with $m$ an integer number of fractional flux quanta).
More intriguingly, however, this vanishing of $\Delta \cS$ occurs for specific values of $q$:
while levitons in the FQH are also minimal excitations, they do not carry a fractional charge and instead correspond to an \emph{integer} number of electrons. 
	This shows that integer levitons are minimal excitation states even in the presence of strong electron-electron interactions, and that it is not possible to excite individual fractional quasiparticles using a properly quantized Lorentzian voltage pulse in time.
	Indeed, it is easy to note that, under these conditions, at $q= \nu$ (single quasiparticle charge pulse)  no specific feature appears in the noise and $\Delta \cS\neq 0$.
	While fractional minimal excitations may exist, they cannot be generated using either Lorentzian, sine or square voltage drives.

Close to integer $q$ the behavior of $\Delta \cS$ is strongly asymmetric. While a slightly larger than integer value leads to vanishingly small excess noise, a slightly lower one produces a seemingly diverging contribution. Indeed, exciting less than a full electronic charge produces a strong disturbance of the ground state, and ultimately leads to the generation of infinitely many particle-hole excitations, which is reminiscent of the orthogonality catastrophe~\cite{levitov96, lee93, anderson67}.

\begin{figure}[tb]
	\includegraphics[width=8.6cm]{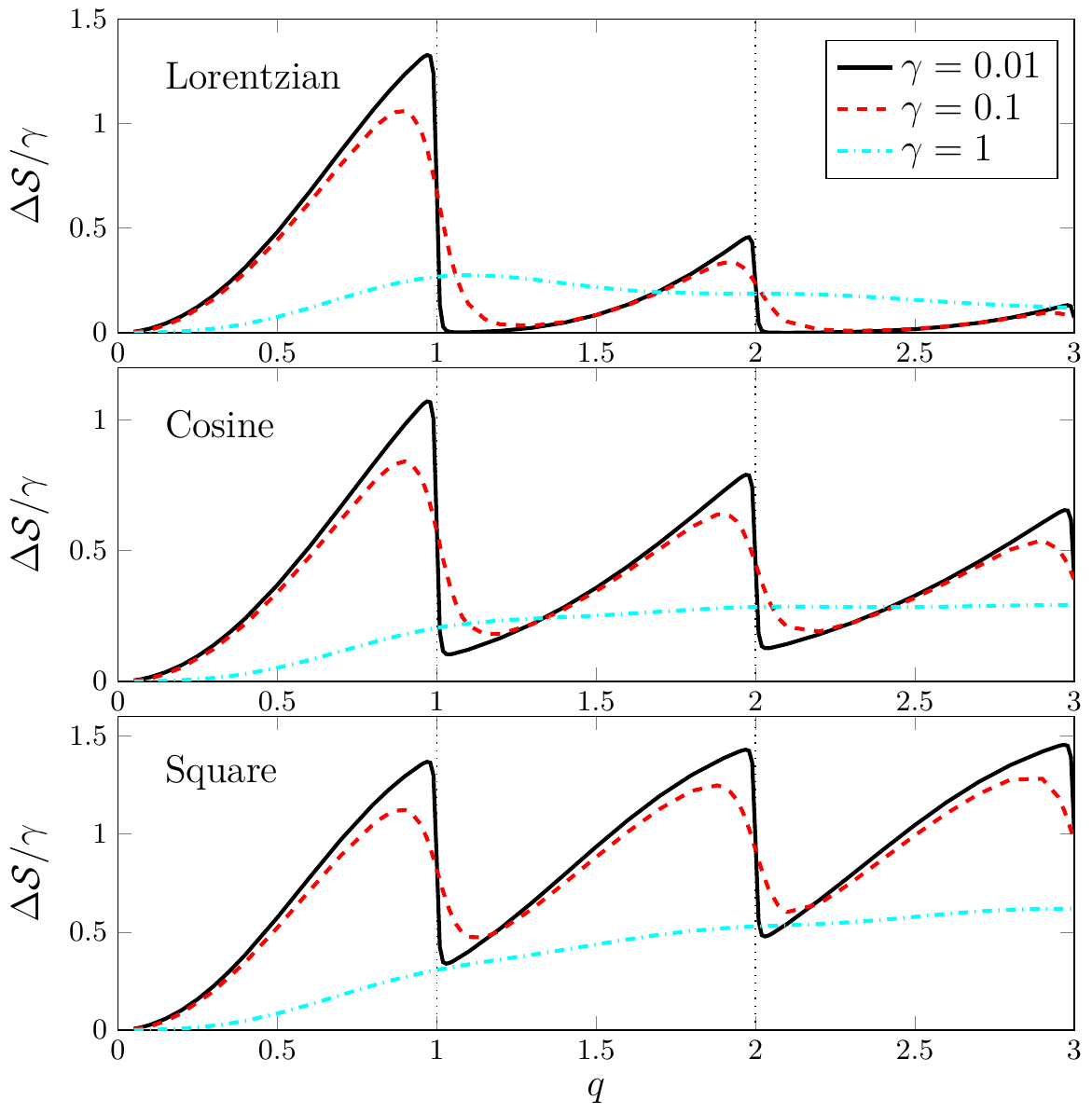}
	\caption{Rescaled excess noise $\Delta \cS / \gamma$ in units of $\frac{e^2}{T}$ as a function of the number of electrons per pulse $q$, for different values of the dimensionless tunneling parameter $\gamma=\frac{|\Gamma_0|^2}{\pi a v_F \Omega}$. Results are obtained at zero temperature, with filling factor $\nu=1/2$, in the case of a square (bottom), a cosine (middle) and a periodic Lorentzian drive with $\eta = 0.1$ (top).}
	\label{fig:xsnoisenuhalf}
\end{figure}

For comparison with experiments, we compute the excess noise at $\theta\neq 0$. This calls for a modified definition of $\Delta \cS$ (in order to discard thermal excitations):
\begin{align}
\Delta \cS = \cS - 2  e^{*} \overline{\langle I_B (t) \rangle} \coth \left( \frac{q}{2 \theta} \right) ,
\end{align}
which coincides with Eq.~\eqref{eq:DeltqSperturb} in the $\theta \to 0$ limit. 
The finite temperature results (see Fig.~\ref{fig:xsnoise}) cure some inherent limitations of the perturbative treatment at $\theta=0$ (diverging behavior close to  integer $q$). The noiseless status of the Lorentzian drive is confirmed, as $\Delta \cS\simeq 0$ at low enough temperature for some values of $q$ (yet shifted compared to the $\theta=0$ ones).

Our perturbative analysis is valid when the differential conductance is smaller than $G_0$. This condition can be achieved on average ($\Gamma_0$ is then bounded from above), but it is not fulfilled in general when the voltage drops near zero because of known divergences at zero temperature. In order to go beyond this WB picture, we now turn to an exact non-perturbative approach for the special filling $\nu=1/2$. While this case does not correspond to an incompressible quantum Hall state, it nevertheless provides important insights concerning the behavior of physical values of $\nu$ beyond the WB regime. The agreement between the two methods in the regime where both are valid makes our results trustworthy. 

We thus extend the refermionization approach for filling factor $\nu=1/2$~\cite{chamon96,sharma03} to a generic AC drive~\cite{Note1}. 
Starting from the full Hamiltonian expressed in terms of bosonic fields, one can now write the tunneling contribution introducing a new fermionic entity, $\psi(x,t) \propto e^{i (\phi_R(x,t) + \phi_L(x,t))/\sqrt{2}}$. Solving the equation of motion for $\psi (x,t)$ near $x=0$, one can define a relation between this new field taken before ($\psi_b$) and after ($\psi_a$) the QPC  
\begin{align}
	\psi_a (t) = \psi_b (t) &- \gamma \Omega  e^{i \varphi(t)+ i q \Omega t} \int_{-\infty}^t dt' e^{- \gamma \Omega (t-t')} \nonumber \\
	& \times \left[ e^{-i \varphi(t') - i q \Omega t'}  \psi_b (t') - \text{H.c.} \right] ,
\end{align}
allowing us to treat the scattering at the QPC at all orders. Expressing the current and noise in terms of $\psi_a$ and $\psi_b$, and using the standard correlation function $\langle \psi_b^\dagger (t) \psi_b(t')\rangle = \int \frac{d\omega}{2\pi v_F} e^{i \omega (t-t')} f(\omega)$ (with $f$ the Fermi function), we derive an exact solution for both the backscattered current and PASN.
As the DC noise at a QPC does not remain Poissonian when its transmission increases, our definition of $\Delta \cS$ is further extended to treat the non-perturbative regime.
In the $\nu=1$ case, where an exact solution exists, it is standard to compare the PASN to its equivalent DC counterpart~\cite{keeling06, dubois13} obtained with the same $V_{dc}$, and $V_{ac}=0$. Here, in order to account for the nontrivial physics involved at the QPC in the FQH, it makes sense to compare our PASN to the DC noise which one obtains for the same charge transferred at the QPC, over one period of the AC drive. At zero temperature, $\Delta \cS$ is redefined as:
\begin{align}
\Delta \cS = \cS  -  2 e^* \overline{\langle I_B \rangle}  +  \frac{\left( e^* \right)^2}{T} 2 \gamma \sin \left(\frac{T}{\gamma e^*} \overline{\langle I_B \rangle} \right) ,
\end{align}
where $e^* = \nu e= e/2$, and $\gamma=\frac{|\Gamma_0|^2}{\pi a v_F \Omega}$ is the dimensionless tunneling parameter.
This definition coincides with the Poissonian one Eq.~\eqref{eq:DeltqSperturb} at low $\gamma$, in that it vanishes for the same values of $q$.

\begin{figure}[tb]
	\includegraphics[width=8.6cm]{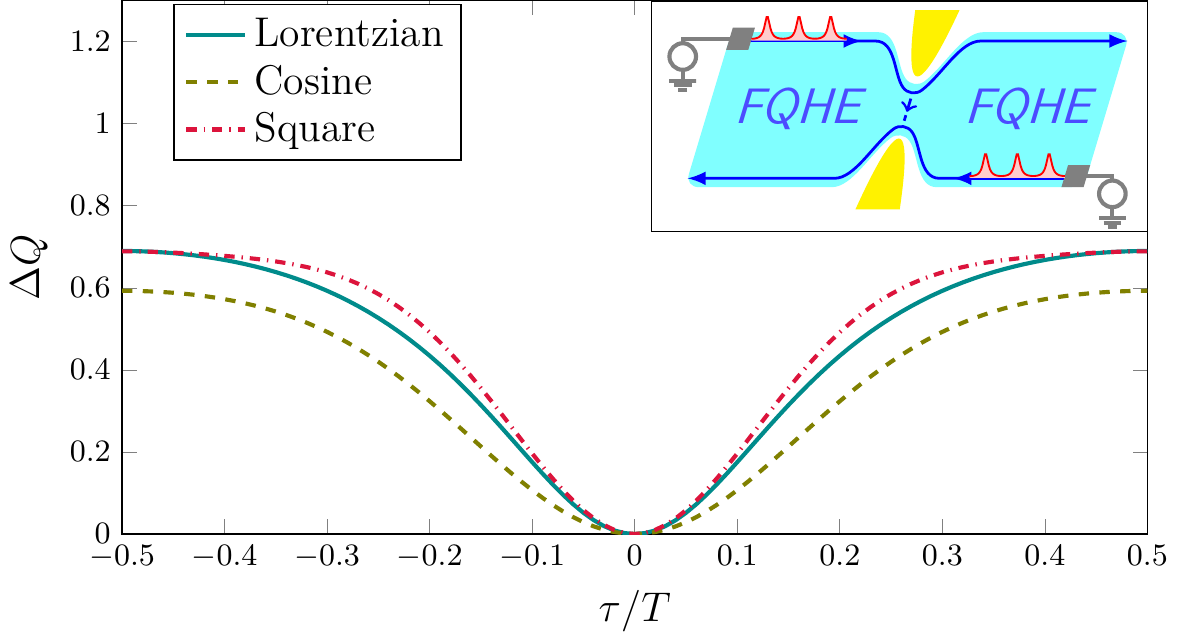}
	\caption{Normalized HOM noise $\Delta Q$ at $q=1$, as a function of the time delay $\tau$ between pulses. Results are presented in the WB case at $\nu=1/3$ and $\theta=0.1$, for a square, a cosine and a periodic Lorentzian drive with $\eta = 0.1$.
		Inset: HOM setup with applied drives on both incoming arms.
	}
	\label{fig:hom}
\end{figure}

Results for $\Delta \cS$ at $\nu=1/2$ are presented in Fig.~\ref{fig:xsnoisenuhalf}.  At low $\gamma$, structures appear as a function of $q$, which are very similar to the perturbative calculation (Fig.~\ref{fig:xsnoise}).
For the Lorentzian drive only, the excess noise approaches zero close to integer values of $q$ in the tunneling regime $\gamma \ll 1$.
When increasing $\gamma$ the position of these minima gets shifted and the excess noise eventually becomes featureless, independently of the AC drive. In the $\gamma \rightarrow +\infty$ limit (not shown) the Lorentzian drive shows signatures of Poissonian electron tunneling at the QPC occurring at $q$ multiples of $\nu$, consistent with the duality property of the FQH regime~\cite{chamon96}.
This Poissonian behavior, not observed for other drives, is also confirmed by the strong backscattering perturbative treatment.
At finite temperature, our results are almost unaffected for $\theta \lesssim \gamma$, while larger temperatures tend to smear any variations in $q$.

Levitons can also be explored in the time domain through electronic Hong-Ou-Mandel (HOM) interferometry~\cite{jullien14, dubois13}. By driving both incoming channels, one can study the collision of synchronized excitations onto a beam-splitter, as two-particle interferences reduce the current noise at the output, leading to a Pauli-like dip.
Fig.~\ref{fig:hom} shows the normalized HOM noise $\Delta Q$~\cite{bocquillon13} as a function of the time delay $\tau$ between applied drives. 
While this does not constitute a diagnosis for minimal excitations, it reveals the special nature of levitons in the WB regime, as the normalized HOM noise is independent of temperature and filling factor~\cite{Note1, glattli16}, reducing at $q=1$ to the universal form:
\begin{align}
\Delta Q (\tau) = \frac{\sin^2 \left( \frac{\pi \tau}{T}\right)  }{\sin^2 \left( \frac{\pi \tau}{T}\right) + \sinh^2 \left( 2 \pi \eta \right)} .
\end{align}
The same universal behavior is also obtained for fractional $q=\nu$ in the strong backscattering regime (tunneling of electrons at the QPC).
Interestingly, although the HOM noise and the PASN are very different from their Fermi liquid counterparts, 
an identical expression for $\Delta Q (\tau)$ was also obtained in this case~\cite{dubois13b} (where it is viewed as the overlap of leviton wavepackets). 

\begin{figure}[tb]
\includegraphics[width=8.6cm]{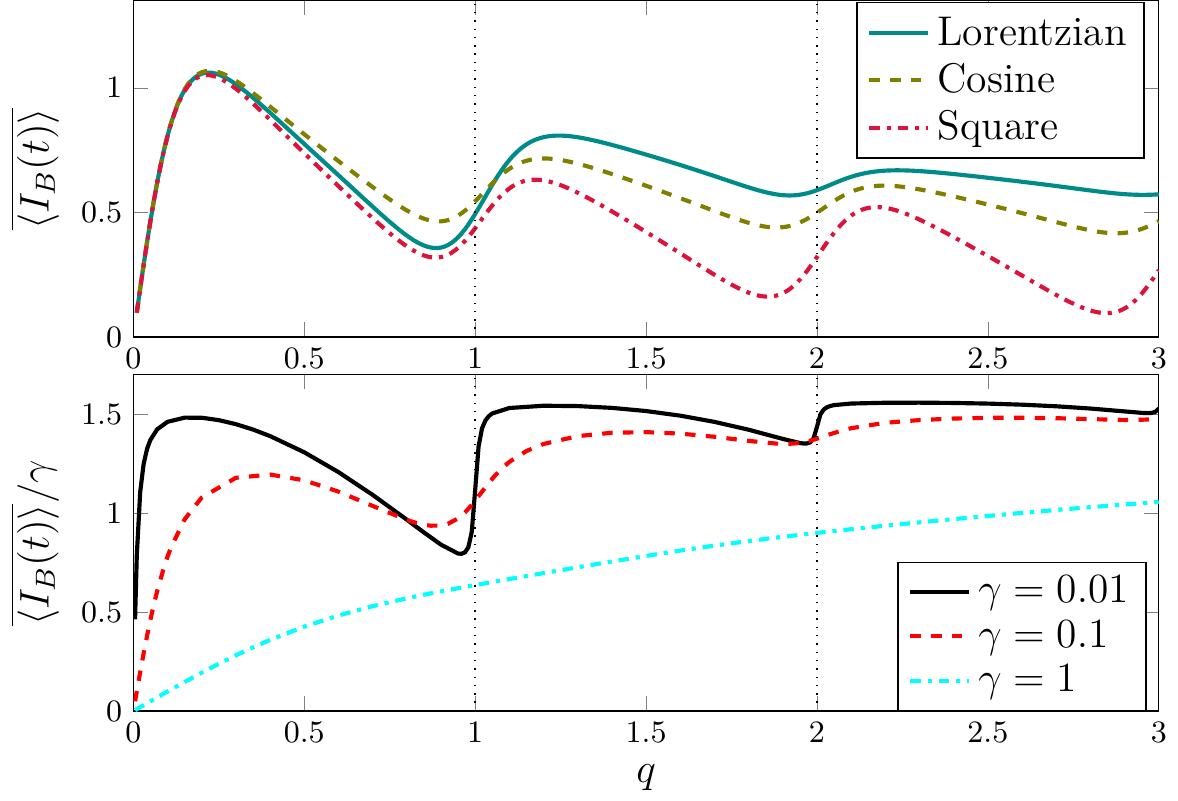}
\caption{Averaged backscattered current $\overline{\langle I_B (t) \rangle}$ as a function of the number of electrons per pulse $q$, in the case of a square, cosine and periodic Lorentzian drive with $\eta = 0.1$ and $\theta=0.1$. Results are presented for $\nu=1/3$ in the perturbative regime in units of $I^0 =\frac{e^{*}}{T} \left( \frac{\Gamma_0}{v_F} \right)^2 \left( \frac{\Omega}{\Lambda} \right)^{2\nu-2}$ (top) and for the exact treatment at $\nu=1/2$ in units of $\frac{e}{T}$ (bottom).}
\label{fig:ibvsq}
\end{figure}

Finally, in addition to the excess noise, the time-averaged backscattering current $\overline{\langle I_B (t) \rangle}$ also bears peculiar features.
In contrast to the Ohmic behavior observed in the Fermi liquid case, $\overline{\langle I_B (t) \rangle}$  shows large dips for integer values of $q$ (see Fig.~\ref{fig:ibvsq}).  These dips are present for all types of periodic drives, and cannot be used to detect minimal excitations.
However, the spacing between these dips provides an alternative diagnosis (from DC shot noise~\cite{depicciotto97, saminadayar97}) to access the fractional charge $e^*$ of Laughlin quasiparticles, as $q$ is known from the drive frequency and the amplitude $V_{dc}$~\cite{crepieux04}.

Real-time quasiparticle wave packet emission has thus been studied in a strongly correlated system,
showing the existence of minimal excitations (levitons) in edge states of the FQH. These occur when applying
a periodic Lorentzian drive with quantized flux, and can be detected as they produce 
Poissonian noise at the output of a Hanbury-Brown and Twiss setup in the weak backscattering regime. Although FQH quasiparticles typically carry a fractional charge, the charge of these noiseless excitations generated through Lorentzian voltage pulse corresponds to an integer number of $e$. Furthermore,
our findings are confirmed for arbitrary tunneling using an exact refermionization scheme. Remarkably enough, in spite of the strong interaction, two FQH leviton collisions bear a universal Hong-Ou-Mandel signature identical to their Fermi liquid analog.   
Possible extensions of this work could address more involved interferometry of minimal excitations as well as their generalization to non-Abelian states.


\textbf{Note added in proofs: } During the completion of this work, it came to our attention that a
simple argument can rule out the possibility of minimal excitations
beyond the results presented here. Starting from Eq.~(7), one readily
sees that a minimal excitation ($\Delta \cS = 0$) can only be realized if
$P_l = 0$ for all $l \leq -q$, independently of the filling factor. At
$\nu=1$, it was shown [7-9] that minimal excitations were
associated with quantized Lorentzian pulses, so that this type of drive
is the only one satisfying the constraint of vanishing $P_l$. Since this
condition is independent of $\nu$, it follows that also at fractional
filling, minimal excitations can only be generated using Lorentzian
drives with quantized charge $q \in \mathbb{Z}$.

\begin{acknowledgments}
We are grateful to  G. F\`eve, B. Pla\c cais, P. Degiovanni and D.C. Glattli for useful discussions. This work was granted access to the HPC resources of Aix-Marseille Universit\'e financed by the project Equip@Meso (Grant No. ANR-10-EQPX-29-01). It  has been carried out in the framework of project "1shot reloaded"
(Grant No. ANR-14-CE32-0017) and benefited from the support of the Labex ARCHIMEDE (Grant No. ANR-11-LABX-0033) and of the AMIDEX project (Grant No. ANR-11-IDEX-0001-02), all funded by the "investissements d'avenir" French Government program managed by the French National Research Agency (ANR).
\end{acknowledgments}

\appendix

\section{External drives and corresponding Floquet coefficients}

The applied drive $V(t)$ is split into a DC and an AC part $V(t) = V_{dc} + V_{ac}(t)$, where by definition $V_{ac}(t)$ averages to zero over one drive period $T$.

The AC voltage is handled through the accumulated phase experienced by the quasiparticles $\varphi (t) = e^{*} \int_{-\infty}^t dt' V_{ac}(t')$ (with the fractional charge $e^* = \nu e$). We use the Fourier decomposition of $e^{-i \varphi(t)}$, defining the corresponding coefficients $p_l$ as
\begin{equation}
p_l = \int_{-T/2}^{T/2} \frac{dt}{T} e^{i l \Omega t} e^{-i \varphi(t)}.
\label{eq:defpofl}
\end{equation}

We focus on three types of drives:
\begin{align}
\text{Cosine} \quad &  V(t) = V_{dc} \left[ 1 - \cos \left( \Omega t \right)\right]  ,\\
\text{Square} \quad &  V(t) = 2 V_{dc} \sum_k \mathrm{rect}\left( 2\frac{t}{T}-2k \right),\\
\text{Lorentzian} \quad &  V(t) = \frac{V_{dc}}{\pi} \sum_k \frac{\eta}{\eta^2 + \left( \frac{t}{T}-k \right)^2},
\end{align}
where $\mathrm{rect}(x)=1$ for $|x|<1/2$ ($=0$ otherwise), is the rectangular function, and $\eta=W/T$ ($W$ is the half-width at half-maximum of the Lorentzian pulse).

The corresponding Fourier coefficients of Eq.~\eqref{eq:defpofl} read, for non-integer $q = \frac{e^* V_{dc}}{\Omega}$:
\begin{align}
\text{Cosine} \quad &  p_l = J_l \left(  -q \right) , \\
\text{Square} \quad &  p_l = \frac{2}{\pi} \frac{q}{l^2-q^2} \sin\left[\frac{\pi}{2} (l-q) \right]  , \\
\text{Lorentzian} \quad &  p_l =  q \sum_{s=0}^{+\infty} \frac{\Gamma (q+l+s)}{\Gamma(q+1-s)} \frac{(-1)^s e^{-2 \pi \eta (2s+l)}}{(l+s)! s!},
\end{align}

\section{Current and noise in the weak backscattering regime}

Fractional quantum Hall (FQH) edges at filling factor $\nu=1/(2n+1)$ are described in terms of a hydrodynamical model~\cite{wen95} through the Hamiltonian of the form 
\begin{align}
\label{eq:fullH}
H &= H_0 + H_V,	\\
\label{eq:H0}
H_0 &= \frac{ v_F}{4 \pi} \sum_{\mu=R,L} \int dx ~ \left( \partial_x \phi_\mu  \right)^2 ,\\
H_V &= - \frac{e \sqrt{\nu}}{2 \pi}  \int dx ~ V(x,t) ~  \partial_x \phi_R
\label{eq:HV}
\end{align}
where we apply a bias $V(x,t)$ which couples to the charge density of the right moving edge state. Here the bosonic fields satisfy $\left[ \phi_{R,L} (x) , \phi_{R,L} (y)\right] = \pm i \pi \text{Sgn} (x-y)$.

These bosonic fields propagate along the edge at velocity $v_{F}$ and are directly related to the corresponding quasiparticle annihilation operators $\psi_\mu (x,t)$ $(\mu=R, L)$ through the bosonization identity  
\begin{equation}
\psi_{R/L} (x,t) = \frac{U_{R/L}}{\sqrt{2 \pi a}} e^{\pm i k_F x} e^{- i \sqrt{\nu} \phi_{R/L} (x,t)},
\label{eq:psiboso}
\end{equation}
where $a$ is a short distance cutoff and $U_\mu$ are Klein factors.

Focusing first on the case where no QPC is present, one can derive the following equations of motion for the bosonic fields 
\begin{align}
\left( \partial_t + v_F \partial_x \right) \phi_R (x,t) &= \frac{e}{\hbar} \sqrt{\nu} V (x,t) \\
\left( \partial_t + v_F \partial_x \right) \phi_L (x,t) &= 0
\end{align}
It follows that the effect of the external voltage bias can be accounted for by a rescaling of the right-moving bosonic field $\phi_{R} (x,t) = \phi_{R}^{(0)} (x,t) + e \sqrt{\nu} \int_{-\infty}^{t} dt' V(x',t') $ (with $\phi^{(0)}_{R}$ the solution in absence of time dependent voltage) or alternatively by a phase shift of the quasiparticle operator of the form
\begin{equation}
\psi_{R} (x,t)  \longrightarrow \psi_{R} (x,t) ~ e^{- i \nu e \int_{-\infty}^t dt' V (x', t')} 
\label{eq:phaseshift}
\end{equation}
where $x' = x - v_F (t-t')$. 

Accounting for this phase shift, the tunneling Hamiltonian which describes the scattering of single quasiparticles at the QPC ($x=0$) in the weak backscattering regime is given by $H_T = \Gamma_0 \exp\left(  i \nu e  \int_{-\infty}^t dt' V (v_F(t'-t), t') \right) \psi_R^\dagger (0) \psi_L (0) +\text{H. c.} $,  with the bare tunneling constant $\Gamma_0$.  

Considering for simplicity the experimentally motivated situation of a long contact,  located at a distance $d$ from the QPC, one can write the bias voltage as $V(x,t) = V(t) 
\theta \left( -x-d \right)$ where $V(t)$ is a periodic time-dependent voltage.
The tunneling Hamiltonian can then be simplified as
\begin{align}
H_T &= \Gamma_0 \exp\left(  i \nu e  \int_{-\infty}^{t - \frac{d}{v_F}} dt'  V(t') 
\right) \psi_R^\dagger (0) \psi_L (0) +H. c. 
\label{eq:HT}
\end{align}
Note that the time delay $d/v_F$ can safely be discarded as it corresponds to a trivial constant shift in time of the external drive.

The backscattering current is readily obtained from $H_T$, after defining $\Gamma (t) = \Gamma_0  \exp\left(  i e^{*}  \int_{-\infty}^t dt' V (t') \right)$ and $e^* = \nu e$: 
\begin{equation}
I_B (t) = i e^{*}  \left[ \Gamma (t) \psi_R^\dagger (0,t) \psi_L (0,t)  - \text{H.c.} \right] .
\end{equation}
Expanding to order $\Gamma_0^2$ and taking the average over one period, the backscattering current becomes
\begin{align}
\overline{\langle I_B (t) \rangle} =&
-
\frac{2 i e^{*}}{T}  \left( \frac{\Gamma_0}{2 \pi a} \right)^2 
 \int_{-\infty}^{+\infty} d\tau e^{2 \nu \cG \left( -\tau \right) } \nonumber \\
& \times \int_0^T d\bart  \sin \left[ e^{*} \int_{\bart-\frac{\tau}{2}}^{\bart+\frac{\tau}{2}} dt'' V (t'') \right],
\end{align}
with  $\cG(t-t')$, the connected bosonic Green's function $\cG(t-t') = \langle \phi_\mu (0,t) \phi_\mu (0,t') \rangle_{c}$, which reads at zero temperature $\cG(t-t') = - \log \left[ 1 +i \frac{v_F(t-t')}{a} \right]$. 

The unsymmetrized shot noise is written in terms of $I_B (t)$~\cite{crepieux04} as $
S (t,t') = \langle I_{B} (t) I_{B} (t') \rangle - \langle I_{B} (t) \rangle \langle I_{B} (t') \rangle$.
To second order $\Gamma_0^2$, the zero-frequency time-averaged shot noise becomes:
\begin{align}
\cS &= 2 \int d\tau \int_0^T \frac{d \bart}{T} S \left( \bart + \frac{\tau}{2}; \bart-\frac{\tau}{2}\right)  \nonumber \\
&= \left( \frac{e^{*} \Gamma_0}{\pi a} \right)^2  \int d\tau e^{2 \nu \cG \left( -\tau \right) } \nonumber \\
& \qquad \times \int_0^T \frac{d \bart}{T} \cos \left[ e^{*} \int_{\bart - \frac{\tau}{2}}^{\bart + \frac{\tau}{2}} dt'' V(t'') \right] .
\label{eq:defPASN}
\end{align}
The excess shot noise at zero temperature then takes the form
\begin{align}
\Delta \cS &= \cS - 2 e^{*} \overline{\langle I_B (t) \rangle} \nonumber \\
&= \left( \frac{e^{*} \Gamma_0}{\pi a} \right)^2  \int d\tau  \int_0^T \frac{d \bart}{T}  \exp \left[2 \nu \cG \left( -\tau\right)  \right] \nonumber \\ 
& \qquad \times \exp \left[ i e^{*} \int_{\bart - \frac{\tau}{2}}^{\bart + \frac{\tau}{2}} dt'' V(t'') \right] .
\label{eq:DeltaSzeroT}
\end{align}
Splitting the voltage into its DC and AC part, and using the Fourier coefficients $p_l$ introduced in Eq.~\eqref{eq:defpofl}, one has
\begin{align}
\Delta \cS &=  \left( \frac{e^{*} \Gamma_0}{\pi a} \right)^2   \sum_{l} |p_l|^2 \int d\tau
e^{ i ( l +q ) \Omega \tau + 2 \nu \cG \left( - \tau \right) }  \nonumber \\
&= 
\frac{2}{T} \left( \frac{e^{*} \Gamma_0}{ v_F} \right)^2
\frac{1}{\Gamma (2 \nu)} \left(\frac{\Omega}{\Lambda}\right)^{2\nu-2} \nonumber \\
& \qquad \times \sum_{l} P_l  \left|  l + q \right|^{2\nu-1}  \left[ 1 - \Sgn \left( l + q \right) \right] ,
\end{align}
where the zero-temperature expression of $\cG(-\tau)$ has been used, which allows to perform the integration. As in the main text, we also introduced the notations $q=\frac{e^{*} V_{dc}}{\Omega}$ for the charge per pulse, $\Lambda = v_F/a$ for the high-energy cutoff of the chiral Luttinger liquid theory and $P_l = |p_l|^2$ for the probability for a quasiparticle to absorb ($l>0$) or emit ($l<0$) $l$ photons~\cite{dubois13b}.

At finite temperature, $\Delta \cS$ needs to be slightly amended,
\begin{align}
\Delta \cS &= \cS - 2 e^{*} \overline{\langle I_B (t) \rangle} \coth \left( \frac{q}{2 \theta} \right) \nonumber \\
&= - \frac{2}{T} \left( \frac{e^{*} \Gamma_0}{v_F} \right)^2  \frac{2  \theta}{\Gamma (2 \nu)}  \left( \frac{2 \pi \Omega\theta}{\Lambda} \right)^{2 \nu-2}  \nonumber \\
& \qquad \times \sum_{l} P_l \left| \Gamma \left( \nu + i \frac{l+ q}{2 \pi \theta} \right) \right|^2  \frac{\sinh \left( \frac{l}{2 \theta} \right)}{\sinh \left( \frac{q}{2 \theta} \right)} ,
\label{eq:DeltaSfiniteT}
\end{align}
in order to get rid of the thermal noise ($\Delta \cS \to 0$ for large temperature). There, the reduced temperature is $\theta = \frac{k_B \Theta}{\Omega}$ ($\Theta$ the electron temperature) and the finite-temperature expression of $\cG \left( - \tau \right)$ has been employed.

\section{Excitation number and signatures of minimal excitations}

The number of excitations created in a one-dimensional system of free fermions by the applied time-dependent drive $V(t)$ is given by the number of electrons and holes
\begin{align}
N_e &= \sum_k n_F(-k) \langle \psi_k^\dagger \psi_k \rangle ,\\
N_h &=  \sum_k n_F(k) \langle \psi_k \psi_k^\dagger \rangle ,
\end{align}
where $n_F(k)$ is the Fermi distribution and $\psi_k$ is a fermionic annihilation operator in momentum space. Using the bosonized description for $\nu=1$ [see Eq.~\eqref{eq:psiboso}], the number of electrons and holes becomes
\begin{align}
N_{e/h} = v_F^2 \int \frac{dt dt'}{(2 \pi a)^2} \exp \left[2 \cG(t'-t) \mp i e \int_{t'}^t d\tau V(\tau) \right] .
\end{align}

Minimal excitations correpond to a drive which excites a single electron, while no particle-hole pairs are generated ($N_h=0$). Generalizing this to a chiral Luttinger liquid (a FQH edge state)~\cite{wen95}, this excitation should correspond to a vanishingly small value of the quantity
\begin{align}
\cN = v_F^2  \int \frac{dt dt'}{(2 \pi a)^2} \exp \left[2 \nu \cG(t'-t) + i e^{*} \int_{t'}^t d\tau V(\tau) \right] .
\end{align}
The excess noise [defined in Eq.~\eqref{eq:DeltaSzeroT}] is thus identified as the most suitable quantity to study minimal excitations.
At $\theta=0$, one recovers in $\cN$ precisely the excess noise, up to a prefactor which depends on the tunneling amplitude [see Eq.~\eqref{eq:DeltaSzeroT}]~\cite{keeling06}. At $\theta\neq 0$, thermal excitations contribute substantially to $\cN$, motivating us to include a correction [the $\coth$ factor in  Eq.~\eqref{eq:DeltaSfiniteT}] which gets rid of this spurious contribution in the high temperature limit.

\section{Non-perturbative treatment at $\nu=1/2$}

The Hamiltonian is identical to that of Eqs.~\eqref{eq:H0}-\eqref{eq:HT}. Introducing new bosonic fields $\phi_\pm = \frac{\phi_R \mp \phi_L}{\sqrt{2}}$, it becomes:
\begin{align}
H_0 &= \frac{ v_F}{4 \pi} \sum_{r=\pm} \int dx \left( \partial_x \phi_r  \right)^2 ,\\
H_T &= \Gamma (t) \frac{e^{i \phi_- (0)}}{2 \pi a} + \Gamma^* (t) \frac{e^{-i \phi_- (0)}}{2 \pi a} .
\end{align}
This form of $H_T$ (specific to $\nu=1/2$) allows us to refermionize the $e^{i \phi_-}$ field, and to decouple $\phi_{+}$, following Ref.~\cite{chamon96}. A new fermionic field $\psi (x)$ and a Majorana fermion field $f$, which satisfy $\left\{ f, f \right\} = 2$ and $\left\{ f, \psi(x)\right\} = 0$, are introduced. These obey the equations of motion:
\begin{align}
-i \partial_t \psi (x,t) =& i v_F \partial_x \psi (x,t) + \frac{\Gamma (t)}{\sqrt{2 \pi a}} f(t) \delta(x)  ,\\ 
-i \partial_t f(t) =& 2 \frac{1}{\sqrt{2 \pi a}} \left[ 
 \Gamma^*(t) \psi(0,t)
-  \text{H.c.}
\right] .
\end{align}

Solving this set of equations near the position $x=0$ of the quantum point contact (QPC), one can relate the fields $\psi_b$ and $\psi_a$ corresponding to the new fermionic field taken respectively before and after the QPC:
\begin{align}
\psi_a (t) = &\psi_b (t) - \gamma \Omega  e^{i \varphi(t)+ i q \Omega t} \int_{-\infty}^t dt' e^{- \gamma \Omega (t-t')} \nonumber \\
& \qquad \qquad  \times \left[ e^{-i \varphi(t') - i q \Omega t'}  \psi_b (t') - \text{H.c.} \right] .
\end{align}

The backscattered current is the difference of the left-moving current after and before the QPC:
\begin{align}
I_B (t) &= \frac{e v_F}{2} \left[ \psi_b^\dagger(t) \psi_b(t) - \psi_a^\dagger(t) \psi_a(t) \right] .
\end{align}
After some algebra, the time-averaged backscattered current becomes:
\begin{align}
\overline{\langle I_B \rangle} &= - \frac{e}{T} \gamma  \sum_l P_l~ \text{Im} \left[ \Psi \left( \frac{1}{2} + \frac{\gamma -i (q+l)}{2 \pi \theta} \right) \right] ,
\end{align}
where $\Psi(z)$ is the digamma function, and the dimensionless tunneling parameter is $\gamma =\frac{|\Gamma_0|^2}{\pi a v_F \Omega}$.

Similarly, the zero-frequency time-averaged shot noise defined in Eq.~\eqref{eq:defPASN} takes the form:
\begin{align}
\cS = &\frac{e^2}{T} 4 \gamma^2 \sum_{klm} \frac{\text{Re}\left(p_k^* p_l p_{l+m}^* p_{k+m}\right)}{ m^2+4\gamma^2 } \nonumber \\
& \times \text{Re} \left[
\left( \frac{\frac{2\gamma^2}{m} - i \gamma}{\tanh \left( \frac{l-k}{2\theta} \right)} 
- \frac{m +i \gamma +\frac{2\gamma^2}{m}}{\tanh \left( \frac{k+l+m+2q}{2\theta} \right)} \right) \right. \nonumber \\
& \qquad  \qquad \times  \Psi \left( \frac{1}{2} + \frac{\gamma -i (k+q)}{2\pi \theta}\right) 
\Bigg] .
\label{eq:PASNnuhalf}
\end{align}

Finally, the excess shot noise is obtained using the known $\theta=0$ DC results~\cite{chamon96} for the backscattered current $\langle I_B \rangle_{dc}$ and corresponding zero-frequency noise $\cS_{dc}$:
\begin{align}
\langle I_B \rangle_{dc} &= \frac{e}{2 \pi} \xi \arctan \left(\frac{e V_{dc}}{2 \xi}\right) , \\
\cS_{dc} &= \frac{e^2}{2 \pi} \xi \left[ \arctan \left(\frac{e V_{dc}}{2 \xi}\right)  - \frac{\frac{e V_{dc}}{2 \xi}}{1+\left( \frac{e V_{dc}}{2 \xi}\right)^2} \right] ,
\end{align}
where we introduced $\xi = \frac{|\Gamma_0|^2}{\pi a v_F}$. This allows to express the DC noise, not as a function of the applied bias, but rather as a function of the charge $\cQ_{\Delta t} = \Delta t \langle I_B \rangle_{dc}$ transferred through the QPC over a given time interval $\Delta t$
\begin{align}
\cS_{dc} (\cQ_{\Delta t}) = e \frac{\cQ_{\Delta t}}{\Delta t} - \frac{e^2 \xi }{4 \pi} \sin \left( \frac{4 \pi}{\xi e} \frac{\cQ_{\Delta t}}{\Delta t} \right) .
\end{align}

The excess noise at $\theta=0$  associated with an arbitrary drive $V(t)$ is then defined as the difference between the photo-assisted shot noise (PASN) and the DC noise $\cS_{dc} (\cQ_T)$ obtained for the same charge $\cQ_T = T \overline{\langle I_B \rangle} $ transferred during one period of the AC drive
\begin{align}
\Delta \cS = \cS  -  2 e^* \overline{\langle I_B \rangle}  +  \frac{\left( e^* \right)^2}{T} 2 \gamma \sin \left(\frac{T}{\gamma e^*} \overline{\langle I_B \rangle} \right) , 
\end{align}
where we reintroduced the effective charge $e^* = e/2$.

\section{Hong-Ou-Mandel collision of levitons}

A periodic voltage bias is now applied to both right- and left-moving incoming arms of the QPC. We focus here on single leviton collisions with identical potential drives, up to a tunable time delay $\tau$. The drives are periodic Lorentzians with a single electron charge per pulse ($q_R = q_L = 1$). Using a gauge transformation, this amounts to computing the noise in the case of a single total drive $V_\text{Tot} (t) = V(t) - V(t-\tau)$ applied to the right incoming branch only.

In the context of electron quantum optics, a standard procedure is to compare this so-called Hong-Ou-Mandel (HOM) noise to the Hanbury-Brown and Twiss (HBT) case where single levitons scatter on the QPC without interfering, leading to the definition of the following normalized HOM noise~\cite{bocquillon13,bocquillon14}
\begin{align}
\Delta Q (\tau) = \frac{\cS_{V(t)-V(t-\tau)}  - \cS_\text{vac}}{\cS_{V(t)} + \cS_{V(t-\tau)} - 2 \cS_\text{vac}} .
\label{eq:DeltaQ}
\end{align}
Thermal fluctuations are eliminated by subtracting the vacuum contribution $\cS_{\text{vac}}$ to each instance of the noise.

Taking advantage of the gauge transformation, we use the expressions for the noise established earlier in the perturbative and the exact cases. This calls for a new set of Fourier coefficients $\tilde{p}_l$ associated with the total drive $V_\text{Tot} (t)$:
\begin{align}
\tilde{p}_l =& \sum_m p_{m} p_{m-l}^* e^{i (l-m) \Omega \tau} \nonumber \\
=& \frac{2 i \sin \left(\pi \frac{\tau}{T}\right) e^{i (l+1)\pi \tau/T} (1-z^2)}{1- z^2 e^{-2 i \pi \tau/T}} \nonumber \\
& \times \left[
\left( \frac{1}{e^{2 i \pi \tau/T} -1} + \frac{1}{1-z^2} \right) \delta_{l,0} - \left( z e^{-i \pi \tau/T}\right)^{|l|}
\right] ,
\end{align}
where $z = e^{-2 \pi \eta}$ and the Fourier coefficients $p_l$ corresponding to a periodic Lorentzian drive $V(t)$ with $q=1$ take the form
\begin{align}
p_l &= \theta_H \left( l+1^- \right)  z^l \left( 1 - z^2 \right) - \delta_{l,-1} z ,
\label{eq:plq1}
\end{align}
with $\theta_H (x)$ the Heaviside step function.

In the WB case, the PASN is given by Eq.~\eqref{eq:defPASN}, which gives at finite temperature
\begin{align}
\cS_{V_\text{Tot} (t)} =& S^0 \frac{ \left( 2 \pi \theta \right)^{2 \nu-1}}{\pi \Gamma (2 \nu)}  \sum_{l} |\tilde{p}_l (\tau)|^2   \nonumber \\
& \qquad \times \left| \Gamma \left( \nu + i \frac{l}{2 \pi \theta} \right) \right|^2 \cosh \left( \frac{l}{2 \theta} \right) , \\
\cS_{V (t)} =& S^0 \frac{ \left( 2 \pi \theta \right)^{2 \nu-1}}{\pi \Gamma (2 \nu)}   \sum_{l} |p_l|^2  \nonumber \\
& \qquad \times \left| \Gamma \left( \nu + i \frac{l+q}{2 \pi \theta} \right) \right|^2 \cosh \left( \frac{l+q}{2 \theta} \right) ,
\end{align}
while the vacuum contribution reduces to
\begin{align}
\cS_\text{vac} = S^0 \frac{ \left( 2 \pi \theta \right)^{2 \nu-1}}{\pi \Gamma (2 \nu)}   \left[ \Gamma \left( \nu  \right) \right]^2  .
\label{eq:Svac}
\end{align}

Combining the results from Eqs.~\eqref{eq:DeltaQ} through \eqref{eq:Svac}, we obtain for the normalized HOM noise
\begin{align}
\Delta Q (\tau) = \frac{\sin^2 \left( \frac{\pi \tau}{T}\right)  }{\sin^2 \left( \frac{\pi \tau}{T}\right) + \sinh^2 \left( 2 \pi \eta \right)} . 
\label{eq:HOMq1}
\end{align}
Remarkably, this result is independent of both temperature and filling factor. Indeed, thermal contributions factorize in the exact same way in the numerator and denominator, leading to a universal profile. This result also corresponds to that of Ref.~\cite{dubois13} for $\nu=1$.

\section{Applying the voltage bias to a point-like or a long contact}

In the main text, we focus on the experimentally relevant case of a long contact~\cite{dubois13} where electrons travel a long way through ohmic contacts before reaching the mesoscopic conductor, accumulating a phase shift along the way. In a previous work~\cite{keeling06} however, the authors consider applying the voltage pulse through a point-like contact. Here we show that these two approaches are equivalent.

Indeed, one can see starting from the Hamiltonian \eqref{eq:fullH}, and solving the corresponding set of equations of motion for the fields that the external bias can be accounted for by implementing a phase shift of the quasiparticle operator, which we recall here
\begin{equation}
\psi_{R} (x,t)  \longrightarrow \psi_{R} (x,t) ~ e^{- i \nu e \int_{-\infty}^t dt' V (x - v_F (t-t'), t')}. 
\end{equation}

Several choices for the external drive $V(x,t)$ are thus acceptable, provided that the integral $\int_{-\infty}^t dt' V (x - v_F (t-t'), t')$ leads to the same result. Indeed, this phase shift is the only meaningful physical quantity, which gives us some freedom in the choice of $V(x,t)$. We further consider two different options.

In the present work, we apply the voltage to a long contact, so that $V_1(x,t) = \theta (-x-d) V(t)$. This leads to a phase shift of the form
\begin{align}
\Phi_1 &= \int_{-\infty}^t dt' V_1 (x - v_F (t-t'), t') \nonumber \\
&=  \int_{-\infty}^t dt'~  \theta (-x + v_F (t-t')-d) V(t') \nonumber \\
&=  \int_{-\infty}^{t-\frac{x+d}{v_F}} dt' V ( t').
\end{align}

Following now Ref.~\onlinecite{keeling06}, one can recast the single particle Hamiltonian defined in their Eq.~(4) into a form similar to the one presented here in Eq.~\eqref{eq:HV} provided that one defines the applied voltage drive as $V_2 (x,t) = v_F \delta (x+d) \int_{- \infty}^t d\tau V(\tau)$. This, in turn, leads to the phase shift
\begin{align}
\Phi_2 &= \int_{-\infty}^t dt' V_2 (x - v_F (t-t'), t') \nonumber \\
&= \int_{-\infty}^t dt' v_F \delta (x - v_F (t-t')+d) \int_{- \infty}^{t'} d\tau V(\tau) \nonumber \\
&= \int_{- \infty}^{t - \frac{x+d}{v_F}} d\tau V(\tau)
\end{align}

One thus readily sees that at the level of the phase shift experienced by the quasiparticles as a result of the external drive, the protocol presented in Ref.~\onlinecite{keeling06} and the one presented in the text are completely equivalent.

\bibliography{1shot_biblio}

\end{document}